\documentclass[sigconf]{acmart}
\AtBeginDocument{%
  \providecommand\BibTeX{{%
    \normalfont B\kern-0.5em{\scshape i\kern-0.25em b}\kern-0.8em\TeX}}}

\setcopyright{acmcopyright}
\copyrightyear{2023}
\acmYear{2023}
\setcopyright{acmlicensed}
\acmConference[MM '23] {Proceedings of the 31st ACM International Conference on Multimedia}{October 29--November 3, 2023}{Ottawa, ON, Canada.}
\acmBooktitle{Proceedings of the 31st ACM International Conference on Multimedia (MM '23), October 29--November 3, 2023, Ottawa, ON, Canada}
\acmPrice{15.00}
\acmISBN{979-8-4007-0108-5/23/10}
\acmDOI{	https://doi.org/10.1145/3606038.3616171}

\begin{document}

\title{Automatic Edge Error Judgment in Figure Skating Using 3D Pose Estimation from a Monocular Camera and IMUs}


\author{Ryota Tanaka}
\email{tanaka.ryota@g.sp.m.is.nagoya-u.ac.jp}
\affiliation{%
  \institution{Nagoya University}
  \city{Nagoya}
  \country{Japan}
}

\author{Tomohiro Suzuki}
\email{suzuki.tomohiro@g.sp.m.is.nagoya-u.ac.jp}
\affiliation{%
  \institution{Nagoya University}
  \city{Nagoya}
  \country{Japan}
}

\author{Kazuya Takeda}
\email{kazuya.takeda@nagoya-u.jp}
\affiliation{%
  \institution{Nagoya University}
  \city{Nagoya}
  \country{Japan}
}

\author{Keisuke Fujii}
\email{fujii@i.nagoya-u.ac.jp}
\affiliation{%
  \institution{Nagoya University}
  \city{Nagoya}
  \country{Japan}
}

%%
%% By default, the full list of authors will be used on the page
%% headers. Often, this list is too long and will overlap
% \renewcommand{\shortauthors}{Tanaka, et al.}
\renewcommand{\shortauthors}{Ryota Tanaka, Tomohiro Suzuki, Kazuya Takeda, \& Keisuke Fujii}
%% The abstract is a summary of the work to be presented in the
\begin{abstract}
  Automatic evaluating systems are fundamental issues in sports technologies. In many sports, such as figure skating, automated evaluating methods based on pose estimation have been proposed. However, previous studies have evaluated skaters’ skills in 2D analysis. In this paper, we propose an automatic edge error judgment system with a monocular smartphone camera and inertial sensors, which enable us to analyze 3D motions. Edge error is one of the most significant scoring items and is challenging to automatically judge due to its 3D motion.
  The results show that the model using 3D joint position coordinates estimated from the monocular camera as the input feature had the highest accuracy at $83\%$ for unknown skaters’ data. We also analyzed the detailed motion analysis for edge error judgment. These results indicate that the monocular camera can be used to judge edge errors automatically.
  We will provide the figure skating single Lutz jump dataset, including pre-processed videos and labels at \url{https://github.com/ryota-takedalab/JudgeAI-LutzEdge}. 
\end{abstract}







%% A "teaser" image appears between the author and the affiliation
%  seats. Ichiro Suzuki is preparing to bat.}


%%
%% This command processes the author and affiliation, and title
\maketitle

\section{Introduction}
% スポーツにおいて選手の技術を定量的に評価するのは大事
% CV分野の発展による自動採点システムの貢献の幅
% フィギュアスケートは自動採点化が難しい
Automatic evaluation of athletes’ skills is a fundamental issue in sports technologies.
Based on the advancement of technologies in the field of computer vision, such as human pose estimation, various automatic systems to support people visually evaluating them have been proposed such as physical fitness \cite{Liu+22_Aerobics, Stenum+21, Badiola-Bengoa+21}, martial arts \cite{Hudovernik2022, KUMITRON, Echeverria+21, Nurahmadan+21}, and scoring sports (e.g., gymnastics \cite{Díaz-Pereira+14, Khan+21, Liu+22_Aerobics} and figure skating \cite{Xu+18, xia+22, Delmastro+21, hirosawa+20}).
Among these sports, figure skating has many scoring items, which are critical for performance evaluations, and is challenging in terms of automatic scoring systems due to the limitations of the measurement environment (e.g., a large skating rink) and the dynamism of the movements (e.g., jumps and spins).

% スケートの採点構造の説明
% 先行研究はどんな課題設定がされているか(技の分類、回転不足、総合得点予測)
Figure skating is a scoring sport in which the technical score is awarded for jumps, spins, and steps on the ice, and the program components score is granted for the overall performance presentation, such as expressiveness and harmony with the music. However, it has been pointed out that bias and judging errors due to the subjectivity of judges exist to some extent \cite{Zitzewitz+14, Yukutake+18}, which is a fundamental problem for scoring competitions. For this reason, many previous figure skating studies have proposed automating some of the scoring items, in addition to the end-to-end overall scoring systems \cite{Xu+18, xia+22}. Delmastro et al. proposed to automate the classification of six types of figure skating jump \cite{Delmastro+21}, and Hirosawa and Aoki attempted to automate the under-rotation judging of jumps \cite{hirosawa+20}. Both methods used 2D pose estimation from monocular camera images to solve classification tasks.

%%%%%%%%%%%%%%%%%%%%%%%%%%%%%%%%%%%%%%%%%%%%%%%%%%%%%%%%%%%%%%%%%%%%%%%%%%%%%%%
    \begin{figure}[t]
        \centering
        \includegraphics[scale=0.32]
        {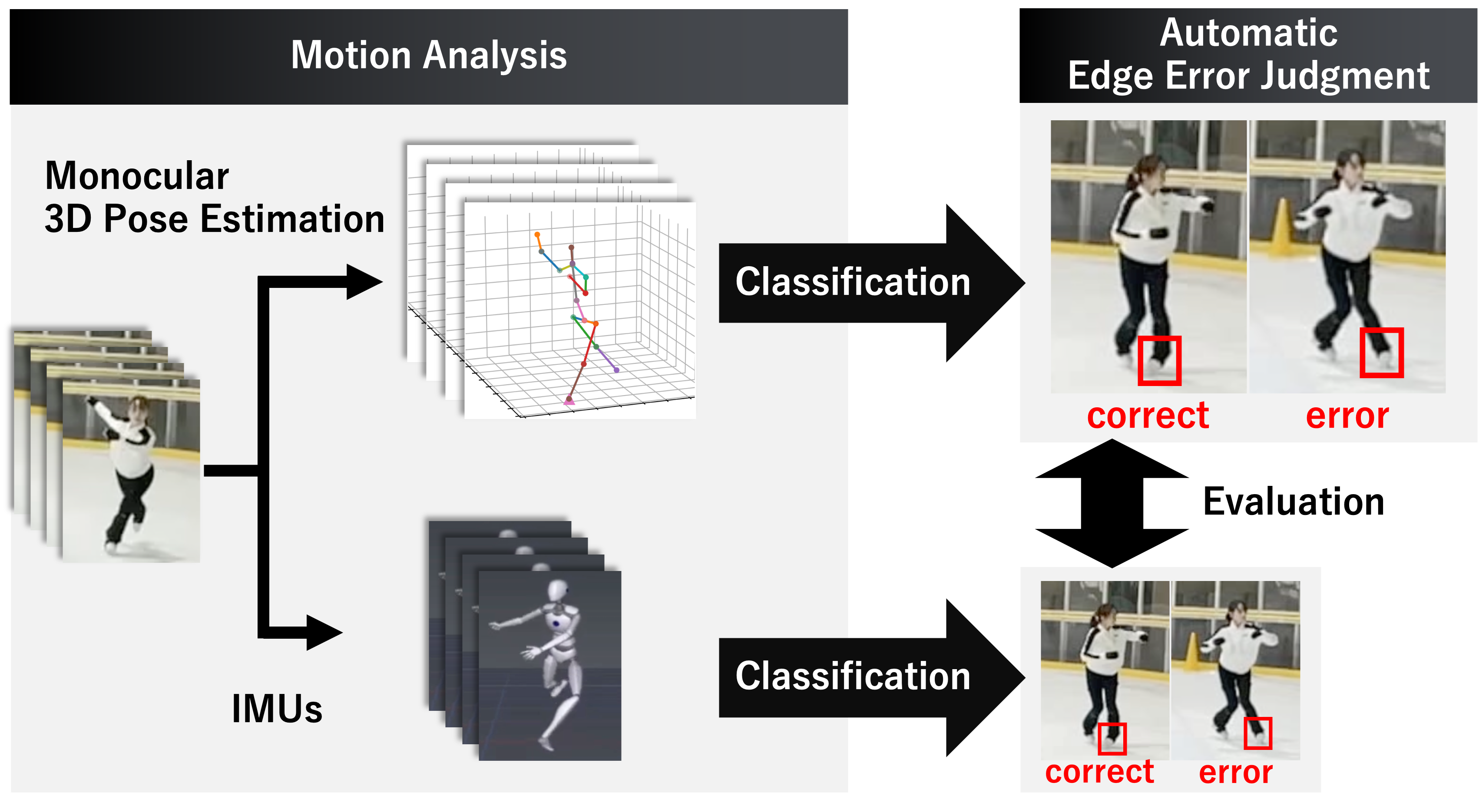}
        \caption{Our system overview. We analyze motions in single Lutz jumps using a monocular camera and IMUs. With the obtained 3D joint position coordinates and joint angles as input features, we separately classify whether the left take-off edges of the attempted jumps are correct or error. Lutz jump is defined as a correct if the left take-off edge is outside and an error if it is inside. Finally, we compare and evaluate what features affected the accuracy of the judgment.
        }
        \label{fig:overview}
    \end{figure}
%%%%%%%%%%%%%%%%%%%%%%%%%%%%%%%%%%%%%%%%%%%%%%%%%%%%%%%%%%%%%%%%%%%%%%%%%%%%%%%

\begin{figure}[h]
        \centering
        \includegraphics[scale=0.25]
        {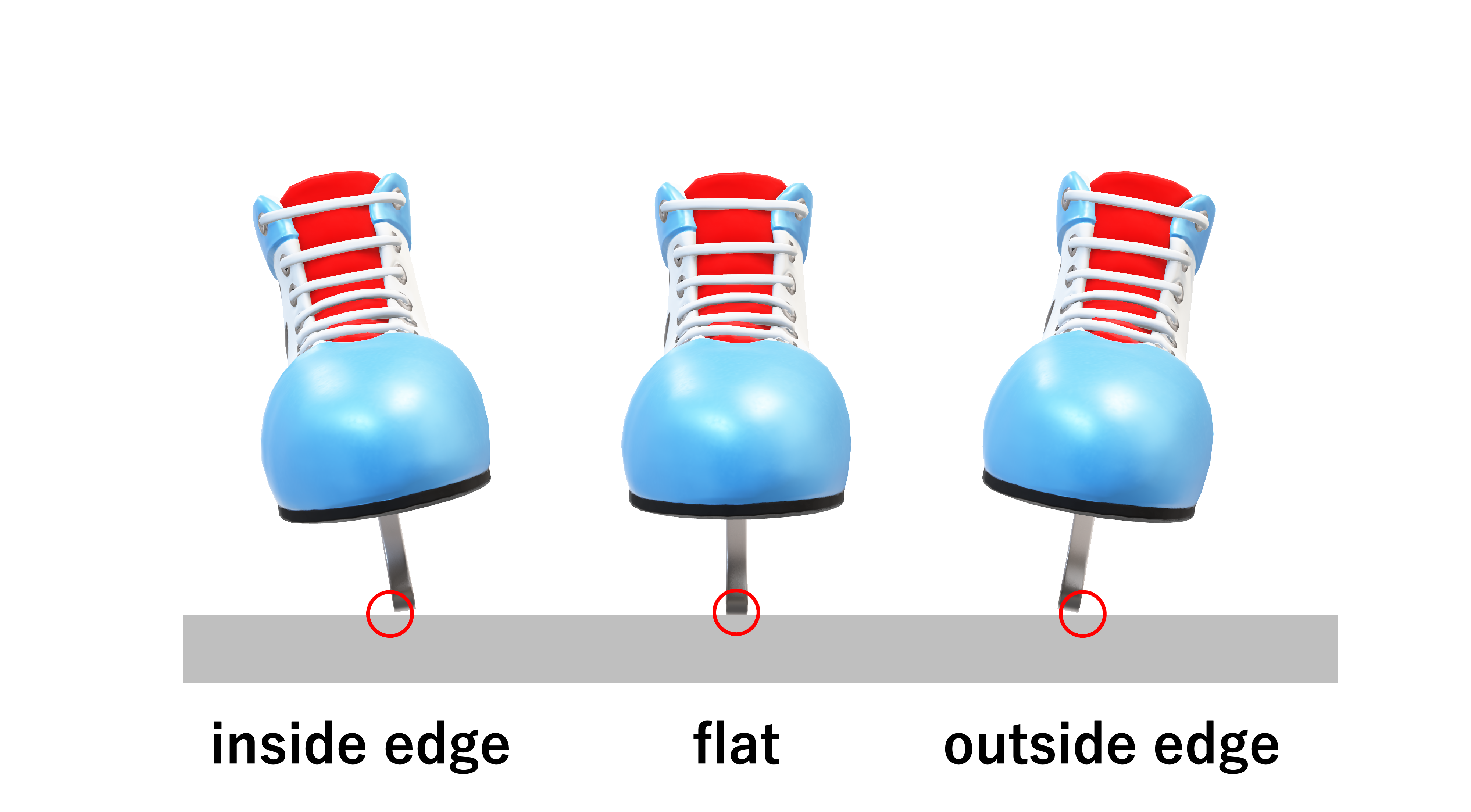}
        \caption{Edge of left foot skate. The inside edge is used by leaning it toward the inside of the body, and the outside edge is used by leaning it toward the outside. The correct Lutz jump is defined as the left skate take-off edge being outside. In other words, the inside edge when the Lutz jump take-off is an error.
        }
        \label{fig:LF_edge}
    \end{figure}

% 先行研究では「エッジ」コントロールの正確さに着目したものがない
% 「エッジ」とは何か
% 「エッジエラー」とは何か
In figure skating, thus, automatic judging jump classification and under-rotation have been researched, but not ``edge'' control accuracy which requires 3D movement analysis. In this paper, we automate edge error judgment as illustrated in Figure \ref{fig:overview}. The term ``edge'' refers to the contact surface between the skate blade and the ice, and the blade has two edges, as shown in Figure \ref{fig:LF_edge}. One of these edges is called the inside edge, which is used with the foot leaning toward the inside of the body, and the other is called the outside edge, which is used with the foot leaning toward the outside of the body.

%%%%%%%%%%%%%%%%%%%%%%%%%%%%%%%%%%%%%%%%%%%%%%%%%%%%%%%%%%%%%%%%%%%%%%%%%%%%%%%
%%%%%%%%%%%%%%%%%%%%%%%%%%%%%%%%%%%%%%%%%%%%%%%%%%%%%%%%%%%%%%%%%%%%%%%%%%%%%%%

Scoring items in figure skating are finely defined, and there are several items in which the accuracy of edge control is related to the addition or subtraction of points. ``Edge error'' is one of these point deduction items to evaluate the accuracy of edge control in figure skating jumps. When a skater jumps with an incorrect edge, it is called an ``edge error''. However, it is often difficult to judge edge errors visually without experience.

Edge errors are prone to occur in the Lutz jump, which is the most difficult among the six types of jumps. The Lutz jump requires using the left skate's outside edge at the take-off, but an edge error occurs when the inside edge is used, as shown in Figure \ref{fig:LF_edge}.
Therefore, the judgment of edge error is significant in competitions because it is related to scoring. However, because of the difficulty of analyzing such minute and 3D movements, research has not yet been conducted to automate this process.

In this paper, we automate edge error judgment using a classification method based on supervised learning from 3D pose estimation from a monocular smartphone camera and Inertial Measurement Units (IMUs) as illustrated in Figure \ref{fig:overview}. 
Motion capture with IMUs has been recently used in figure skating analysis \cite{hanyu+21}, and 3D pose estimation from a monocular camera has been intensively researched \cite{Martinez_2017_ICCV, Güler_2018_CVPR, Pavllo_2019_CVPR, Kocabas_2020_CVPR, StridedTransformer}. 
It has been unknown which 3D pose estimations in IMUs and a monocular camera are better for automatic scoring in figure skating movements.
If they show similar performances, a monocular smartphone camera without sensors is convenient for daily usage in figure skating practices.  
In the experiments, we performed supervised classifications of the edge error for both modalities separately and compared their performance while comparing various inputs of the classifiers.

The contributions of this paper are as follows:
(1) to propose an automatic edge error judgment system using a monocular smartphone camera and IMUs, and to analyze the detailed motion analysis of the pose information contributing to the edge error judgments.
(2) to provide the figure skating single Lutz jump dataset, including pre-processed videos and labels (will be available after publication). 
(3) to demonstrate the effectiveness of automatic judgment using a monocular smartphone camera without sensors by comparing the two types of input sources and various pre-processing types.

\section{Related Work}
% スポーツの自動採点（前半）、自動判定（後半）に関する先行研究を1段落で紹介
% 格闘技やフィットネスの自動採点も入れる。短くて良いので（Slack参照）。
% 水泳（飛び込み）、体操（跳馬）→サッカー（オフサイドは最近出てきた、反則判定も）、陸上（競歩）、ラグビー
\textbf{Automatic scoring system in sports.}
The automatic scoring system in sports has been studied as a part of the Action Quality Assessment \cite{Pirsiavash+14}.
They proposed a system that automatically assesses the quality of actions by regressing scores for an Olympic sports diving and figure skating dataset with spatiotemporal 2D pose features.
Subsequently, Parmar et al. evaluated athletes' skills across various sports \cite{Parmar+19}. Their dataset, which includes seven Olympic events such as diving and gymnastic vault, accelerated research on automated scoring from broadcast videos which was used in \cite{Tang_2020_CVPR, Yu_2021_ICCV, DAE-CoRe_2021, TSA-Net_2022}. However, it may be impractical for competition applications because the evaluation is not based on scoring rules.
For the swimming dive, an automatic scoring system by decomposing the overall score into the execution and difficulty score was proposed \cite{Nekoui+20}. Similarly, for figure skating, which consists of the technical and program components scores, it would be practical to consider the human judges' scoring process to evaluate the athletes' skills.
In rhythmic gymnastics, an automated system assigning judgment scores was proposed \cite{Díaz-Pereira+14} to compare with a set of stored actions scored by an expert judge. 
For other sports, automatic scoring systems have been proposed in physical fitness \cite{Liu+22_Aerobics} with 3D Kinect sensors and martial arts \cite{Hudovernik2022} using 2D human pose estimation.
Compared with prior research aiming to automate the scoring rather than analyzing the basis,  we automate the scoring and analyze the basis of the results by detailed 3D motion analysis.

\noindent \textbf{Foul judgment.}
Foul judgment has also been studied in sports other than scoring competitions. For example, automated offsides detection in soccer \cite{Uchida+21}, automatic detection of faults in race walking \cite{suzuki2022automatic, suzuki2023automatic}, and detection of dangerous tackles in rugby \cite{Nonaka_2022_CVPR} were proposed. These automated judging systems utilized tracking algorithms and human pose estimation to evaluate athletes' skill as a foul detection based on the competition rules. Especially in \cite{suzuki2022automatic, suzuki2023automatic}, they used smartphone cameras to capture athletes' motion and indicated the convenience and performance sufficiency of smartphone cameras as data collection tools.
We also implement automatic judgment using human pose estimation from a smartphone camera as a simpler method to improve availability for many people.

% フィギュアスケートの研究についてまとめて1段落で紹介
% 演技動画を入力とした点数予測
% ジャンプの分類や回転不足の自動判定
% IMUを用いたフィギュアスケート動作解析
% 多視点カメラ映像を用いたフィギュアスケートの動作解析
\noindent \textbf{Scoring in figure skating.}  Research on automatic scoring in figure skating can be broadly divided into two categories: those attempting to predict the overall score \cite{Xu+18, xia+22} and those assistance systems in scoring the technical scores \cite{Liu+20, Delmastro+21, hirosawa+20}. 
Xu et al. proposed the automatic scoring system as a single end-to-end framework \cite{Xu+18}. Still, the scoring basis in this system was unclear, which predicted the overall score without audio information.
Xia et al. automated figure skating overall scoring considering both audio-visual information \cite{xia+22}. However, the scoring basis was still uncertain since they did not score according to the rules defined by International Skating Union (ISU).
To evaluate athletes so that the basis for scoring is more clearly defined, it is desirable to automate judging individual items according to the rules.
Liu et al. proposed an automated system to classify jumps, spins, and steps \cite{Liu+20}.
Delmastro et al. also automated classification for six jump types in figure skating (Axel, Salchow, Toe Loop, Loop, Flip, and Lutz) \cite{Delmastro+21}. In figure skating, the difficulty level varies depending on the type of elements (jumps, spins, and steps) and the scores awarded accordingly. Therefore, automating the classification of elements will assist in scoring.
There also exists a study to judge score deductions for figure skating jumps. Hirosawa et al. proposed an automatic system for determining the under-rotation of jumps using SVM as a classifier after feature extraction using a 3DCNN (C3D) \cite{hirosawa+20}. 
Thus, although the classification of jumps and determination of under-rotation have been studied, no study has evaluated the accuracy of edge control. 

\noindent \textbf{Motion analysis in figure skating.}
Hanyu et al. attempted to analyze figure skating motions using an IMU and showed that the system has good tracking performance for the pose angles of skate blades \cite{hanyu+21}. From the standpoint of convenience and ease of movement, we would not want to attach sensors to the athlete's body, but it could help analyze figure skating jumps to judge edge errors.
Tian et al. proposed a 3D human pose estimation system to analyze figure skating jump motion using six cameras placed outside the skating rink \cite{Tian+20}. The system indicates high estimation quality, but the amount of equipment and labor required to set it up may not be realistic for amateur skaters.
In this paper, we use and compare a monocular smartphone camera and IMUs by automating edge error judgment in figure skating.

\section{Proposed Framework}

% 鈴木くんのCVSportsの構成を真似する感じで、この箇所では、まず最初に本システムの目的や動機を書いて、その後に概要を書く。
% https://ja.overleaf.com/read/bjkxywhcrtwy
In this paper, we propose an automatic edge error judgment system with a monocular smartphone camera and inertial sensors, as shown in Figure \ref{fig:overview}.
Our approach aims to automate figure skating edge error judgment and find the most effective method to judge accurately.
The pose angle of ice skates is necessary to judge edge errors in actual games. However, using optical motion capture with large-scale equipment in an ice skating rink environment is impractical. Therefore, we used IMUs, which can obtain 3D joint position coordinates and the pose angle of ice skates. In addition, we also used 3D pose estimation from monocular camera videos because we ideally do not want to attach sensors to skaters. This method cannot obtain the pose angle of ice skates but 3D joint position coordinates. We evaluated edge error judgment using the pose angles of ice skates obtained by IMUs and the 3D joint position coordinates obtained by IMUs and a monocular camera.

\subsection{3D pose estimation from IMUs}

For data collection by IMUs, we used Perception Neuron 3 (PN3, Noitom, USA). It is equipped with a gyroscope, accelerometer, and magnetometer and can obtain 3D position coordinates and Euler angles of each joint. The previous study suggested that this second-generation model had a high tracking performance for figure skating motion. We attached 17 sensors to participants and captured their movement at 60 fps. We analyzed the motion data with the company's Axis Studio software.

\subsection{3D pose estimation from a monocular video}

For monocular 3D pose estimation, we captured the RGB video data with the 240 fps slow-motion function of the iPhone13 (Apple, USA). To estimate the 3D human pose from the video, it is necessary to deal with the extra people information that appears in the video. Trimming only the frames when the participants jump for the following processing is also desirable. Therefore, we implemented an algorithm that automatically detects only the skater jumping, utilizing YOLOv3 \cite{YOLOv3} and SORT \cite{SORT}: YOLOv3 detects the person in the frame with bounding boxes (BBOXes), and SORT is used to assign a unique ID to each BBOX and track it. The algorithm outputs the BBOX that detected the skater as the one whose vertical velocity change in each BBOX exceeded the threshold value. By identifying the frame when the participant's jump comes to the apex as the moment when the velocity of the BBOX reaches zero, the timing can be aligned and trimmed to a specific number of frames. To estimate the 3D human pose based on the BBOX output with the frames aligned in this way, we utilized StridedTransformer-Pose3D \cite{StridedTransformer} trained on the Human3.6M dataset \cite{human3.6M}. With this 3D human pose estimation model, we can obtain 17 3D joint position coordinates of the whole body.

\subsection{Feature creation}

We obtained 17 3D joint position coordinates from PN3 at 60 fps and StridedTransformer-Pose3D (ST-Pose 3D) \cite{StridedTransformer} at 240 fps. For each of these time series data, we cropped frames so that the take-off moments were aligned. We automated frame cropping using the algorithm described above, especially for the time series data estimated from ST-Pose3D.
The 3D joint position coordinates were then normalized so that the x-y coordinate of the hip and the z-coordinate of the left foot and the right foot closer to the ground were zero. Since the 3D joint position coordinates have 17 points throughout the body and 3 degrees of freedom (x, y, z), which were large feature dimensions, features downsampled to 60 or 12 fps were prepared for comparison. In addition, since PN3 can acquire the pose angle of the shoes, the Euler angles of the movements corresponding to the inside edge and outside edge of the left skate were acquired. Similarly, for the angle information, the features were downsampled to 60 or 12 fps for comparison.

\subsection{Classification}

Using pre-processed data, we implemented edge error judgment by a classification method based on supervised learning. Since the number of data samples was small in this study, we used logistic regression, which can learn with a small number of samples and has the advantage of high interpretability due to the regression coefficients of the model. The model's output is made binary, so the judgment results are 1 for an edge error and 0 for a correct edge.

\section{Experiments}
\subsection{Setup}
\subsubsection{Participants}

Six members of the figure skating club of a university athletic association who hold Level 3 or higher on the badge test (after this, referred to as "skater A-F") participated in this research experiment. The badge test is a proficiency test prescribed by the Japan Skating Federation, and the ranks start from beginner to grades 1-8; the higher the number, the higher the skill level. The single Lutz jump used in this experiment is a task for Level 2 of the badge test, and participants with Level 3 or higher have sufficient skills for this experiment. We provided participants with a thorough explanation of the study and obtained their written consent before the study.

\subsubsection{Measurement}

We attached PN3 inertial sensors to 17 locations on the participant's body to measure the jump data. The IMUs had to be calibrated to obtain 3D joint position coordinates and angles from the acceleration, angular acceleration, and geomagnetic data, hence we calibrated them after attaching them. The data was then simultaneously captured as video and measured by the sensors. 
% The videos were captured with an iPhone 13 on a tripod using the 240 fps slow-motion function. 
We asked participants to attempt edge error jumps intentionally to obtain data on edge error jumps and correct edge jumps. We captured the participants' jumps from the front so that we could record the differences in the edges of each jump. In our experiments, we found that the impact of the jumps caused disturbances in the estimated joint position coordinates and angles. Therefore, we performed one calibration after every subject jumped at least five times.

We excluded the data with clear sensor disturbances or occlusion on video from the data set. The edge errors were judged based on the videos by an official referee with a Class B Technical Specialist (TS) qualification, guaranteed by the Japan Skating Federation. A Technical Specialist is a judge who judges whether a competitor performs elements correctly, including edge error judgment, in figure skating competitions. We utilized the judgment results as the correct answer label for the machine learning model. Table \ref{tab:dataset} shows the number of valid data. In parentheses are the skater's possession levels and years of experience. Skater A and E jumps were all judged as edge errors, but overall we generated a comprehensive data set of 232 with 100 edge errors and 132 clean edges.

%%%%%%%%%%%%%%%%%%%%%%%%%%%%%%%%%%%%%%%%%%%%%%%%%%%%%%%%%%%%%%%%%%%%%%%%%%%%%%%
    \begin{table}[ht]
        \vspace{-5pt}
        \caption{The valid dataset in our experiments.}
        \centering
        \scalebox{1.0}{
        \begin{tabular}{lcc}
            \hline
            & \multicolumn{1}{c}{edge error} & \multicolumn{1}{c}{correct edge} \\
            \hline
            skater A (level 3, 5 years) & $29$ & $0$\\
            skater B (level 3, 4 years) & $19$ & $18$\\
            skater C (level 3, 4 years) & $22$ & $14$\\
            skater D (level 3, 4 years) & $12$ & $38$\\
            skater E (level 3, 5 years) & $30$ & $0$\\
            skater F (level 5, 7 years) & $20$ & $30$\\
            \hline
        \end{tabular}
        }
        \label{tab:dataset}
        \vspace{-10pt}
    \end{table}
%%%%%%%%%%%%%%%%%%%%%%%%%%%%%%%%%%%%%%%%%%%%%%%%%%%%%%%%%%%%%%%%%%%%%%%%%%%%%%%

\subsection{Evaluation}

We used classification models learned from different types of features to compare and evaluate the pre-processed input data. We had two modalities: 3D pose estimation from IMUs and monocular camera videos.
We examined the effectiveness of joint angles and/or 3D joint position coordinates.
As explained in the above sections, we downsampled all data into 60 or 12 fps. 
For this reason, we created the following ten types of features:
\begin{enumerate}
   \item 3D joint pos. coordinates estimated by ST-Pose3D (12 fps)
   \item 3D joint pos. coordinates estimated by ST-Pose3D (60 fps)
   \item 3D joint pos. coordinates captured by PN3 (12 fps)
   \item 3D joint pos. coordinates captured by PN3 (60 fps)
   \item Left skate pose angle captured by PN3 (12 fps)
   \item Left skate pose angle captured by PN3 (60 fps)
   \item Combination of (3) and (5)
   \item Combination of (3) and (6)
   \item Combination of (4) and (5)
   \item Combination of (4) and (6)
\end{enumerate}

The models trained with these features were validated using the player-specific cross-validation method. This method divides the jump data set into training and test data for each skater, cross-validates, and calculates accuracy and F-measure. We calculate accuracy and F-measure as follows, with edge errors as positives and correct edges as negatives.
\begin{displaymath}
  \mathrm{accuracy} = \frac{TP + FP}{TP + TF + FP + FN}
\end{displaymath}
\begin{displaymath}
  \mathrm{F\mathchar`-measure} = \frac{2TP}{2TP + FP + FN}
\end{displaymath}
For example, we set skater A as the test data and the remaining skaters B-F as the training data and judged edge errors by binary classification utilizing logistic regression. This method calculates the average values of accuracy and F-value for each of the six skaters. Since this method does not use the skater data used for test data for model training, it is possible to evaluate the model's generalization performance.

\subsection{Results}

% Classification Results
Table \ref{tab:crossval} shows the classification results: the models using 3D joint position coordinates as input features had the highest accuracy, while the models using the pose angle of the left ice skate as a feature had the lowest accuracy. The model with the highest accuracy used the 3D joint position coordinates predicted by ST-Pose3D as the input feature, with an accuracy of $83.56\%$.
There was little difference in accuracy between downsampled and non-downsampled models, but for the models using features estimated from ST-Pose3D, downsampling resulted in better accuracy. This could be because it prevented over-fitting in the player-specific cross-validation, which requires a generalization ability.

%%%%%%%%%%%%%%%%%%%%%%%%%%%%%%%%%%%%%%%%%%%%%%%%%%%%%%%%%%%%%%%%%%%%%%%%%%%%%%%
    \begin{table}[ht]
        %\vspace{-5pt}
        \caption{The result of player-specific cross-validation.}
        \centering
        \scalebox{0.82}{
        \begin{tabular}{lcc}
            \hline
            & \multicolumn{1}{c}{Accuracy} & \multicolumn{1}{c}{F-Measure} \\
            \hline
            Joint pos. 12fps (ST-Pose3D) & $\bf{83.56\pm13.43\%}$ & $\bf{81.01\pm19.62\%}$\\
            Joint pos. 60fps (ST-Pose3D) & $80.73\pm10.47\%$ & $78.14\pm18.20\%$\\
            Joint pos. 12fps (PN3) & $74.16\pm21.65\%$ & $67.63\pm33.91\%$\\
            Joint pos. 60fps (PN3) & $74.63\pm21.63\%$ & $67.57\pm33.94\%$\\
            Lfoot ang. 12fps (PN3) & $61.09\pm25.28\%$ & $63.98\pm24.54\%$\\
            Lfoot ang. 60fps (PN3) & $58.95\pm26.39\%$ & $64.26\pm21.98\%$\\
            Joint pos. 12fps + Lfoot ang. 12fps (PN3) & $74.60\pm21.10\%$ & $67.80\pm33.89\%$\\
            Joint pos. 12fps + Lfoot ang. 60fps (PN3) & $76.46\pm19.97\%$ & $69.32\pm34.18\%$\\
            Joint pos. 60fps + Lfoot ang. 12fps (PN3) & $75.51\pm20.76\%$ & $68.15\pm33.96\%$\\
            Joint pos. 60fps + Lfoot ang. 60fps (PN3) & $76.89\pm20.48\%$ & $69.16\pm34.25\%$\\
            \hline
        \end{tabular}
        }
        \label{tab:crossval}
        %\vspace{-10pt}
    \end{table}
%%%%%%%%%%%%%%%%%%%%%%%%%%%%%%%%%%%%%%%%%%%%%%%%%%%%%%%%%%%%%%%%%%%%%%%%%%%%%%%

For the model with the highest accuracy, Table \ref{tab:crossval_skaters} shows the results for each skater of the player-specific cross-validation. The table shows that the accuracy of the judgments varied greatly depending on which player's data was used for the training and test data. This considerable variation in judgment results could be attributed to the difference in how each skater jumped the edge error jumps and the correct jumps. This difference in jumping style is discussed later by visualizing the trajectory of the estimated 3D joint position coordinates.

%%%%%%%%%%%%%%%%%%%%%%%%%%%%%%%%%%%%%%%%%%%%%%%%%%%%%%%%%%%%%%%%%%%%%%%%%%%%%%%
    \begin{table}[ht]
        %\vspace{-5pt}
        \caption{The result for skaters (Joint pos. 12fps / ST-Pose3D).}
        \centering
        \scalebox{1}{
        \begin{tabular}{lcc}
            \hline
            & \multicolumn{1}{c}{Accuracy} & \multicolumn{1}{c}{F-Measure} \\
            \hline
            skater A & $100\%$ & $100\%$\\
            skater B & $97.30\%$ & $97.44\%$\\
            skater C & $72.22\%$ & $75.00\%$\\
            skater D & $68.75\%$ & $44.44\%$\\
            skater E & $93.10\%$ & $96.43\%$\\
            skater F & $70.00\%$ & $72.73\%$\\
            \hline
        \end{tabular}
        }
        \label{tab:crossval_skaters}
        %\vspace{-10pt}
    \end{table}
%%%%%%%%%%%%%%%%%%%%%%%%%%%%%%%%%%%%%%%%%%%%%%%%%%%%%%%%%%%%%%%%%%%%%%%%%%%%%%%

\begin{figure}[h]
        \centering
        \includegraphics[scale=0.45]
        {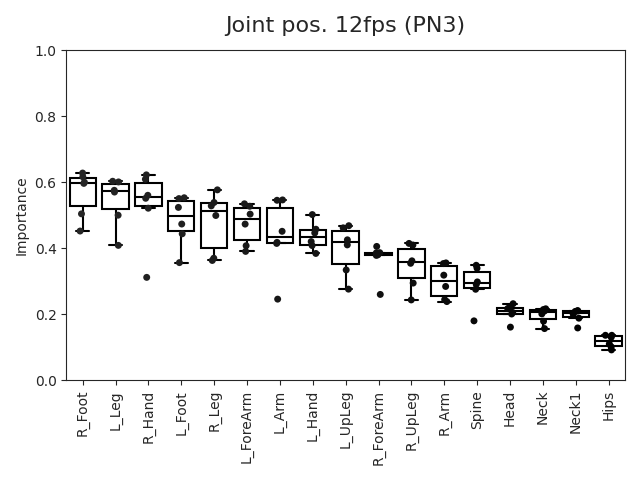}
        \includegraphics[scale=0.45]
        {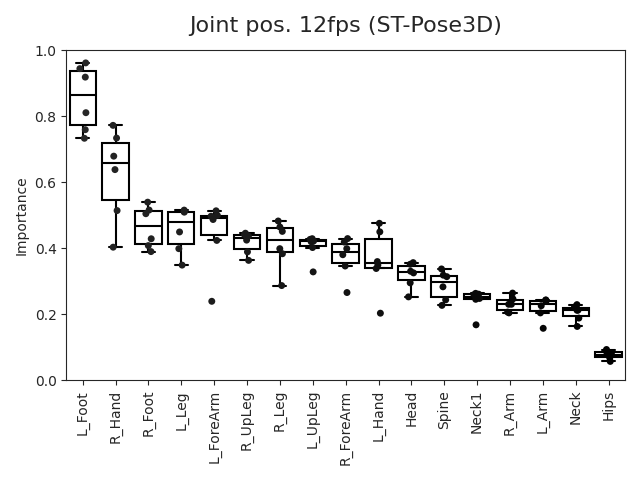}
        \caption{The feature importance in the two models. The model using the 3D joint position coordinates estimated by ST-Pose3D had a larger feature importance for the left foot than the others.}
        \label{fig:coef}
    \end{figure}

% Regression Coefficients
Figure \ref{fig:coef} shows the results of boxplots of the feature importance for each model that used 3D joint position coordinates. The importance was calculated as the average of the absolute values of the regression coefficients at each frame for each feature. There were no notable differences in the magnitude of the importance between features of the model using 3D joint position coordinates captured by PN3. On the other hand, the boxplots show that the model using the 3D joint position coordinates estimated by ST-Pose3D had a larger importance for the left foot than the others. In actual games, referees judge edge errors by looking at the tilt of the left ice skate blade. Although this model did not use the left foot pose angle as a feature, the importance suggests that the model used the relative difference in the left foot position coordinates as an essential feature for judgment.

%%%%%%%%%%%%%%%%%%%%%%%%%%%%%%%%%%%%%%%%%%%%%%%%%%%%%%%%%%%%%%%%%%%%%%%%%%%%%%%
%%%%%%%%%%%%%%%%%%%%%%%%%%%%%%%%%%%%%%%%%%%%%%%%%%%%%%%%%%%%%%%%%%%%%%%%%%%%%%%

\begin{figure}[h]
        \centering
        \includegraphics[scale=0.34]
        {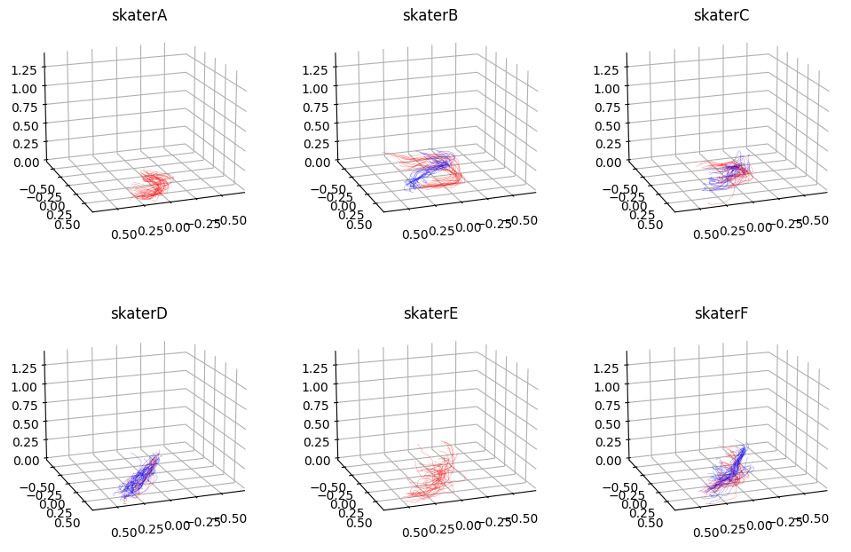}
        \caption{The trajectories of the left foot for six skaters estimated by ST-Pose3D.  A red line represents the trajectory of a Lutz jump judged as an edge error, and a blue line represents the trajectory of a Lutz jump judged as a correct edge. For skaters B, C, and F, the red lines show rounded outward bulging trajectories, whereas the blue lines show straight trajectories.}
        \label{fig:LF_STpose3D}
    \end{figure}

\begin{figure}[h]
        \centering
        \includegraphics[scale=0.45]
        {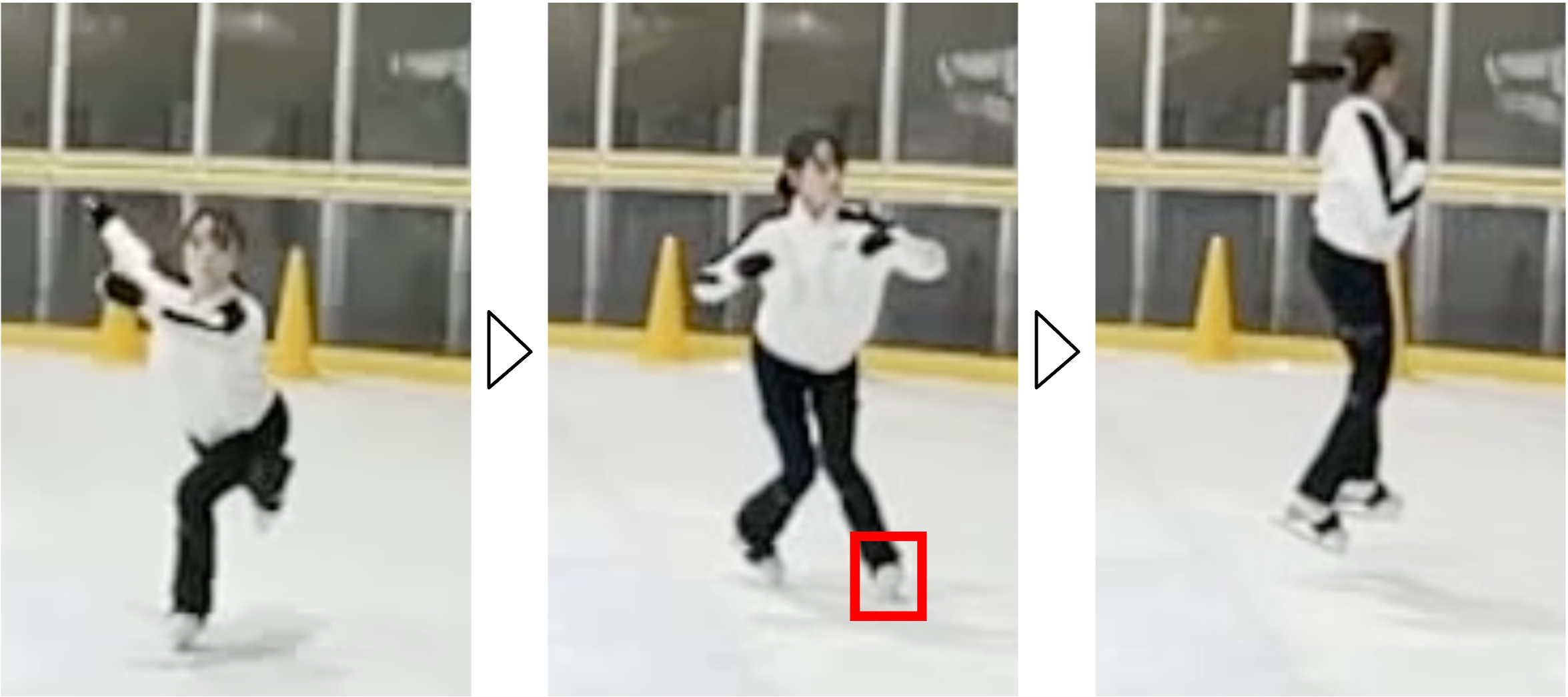}
        \caption{Image examples of take-off with an error edge (inside edge) from the captured video.}
        \label{fig:error}
    \end{figure}

\begin{figure}[h]
        \centering
        \includegraphics[scale=0.45]
        {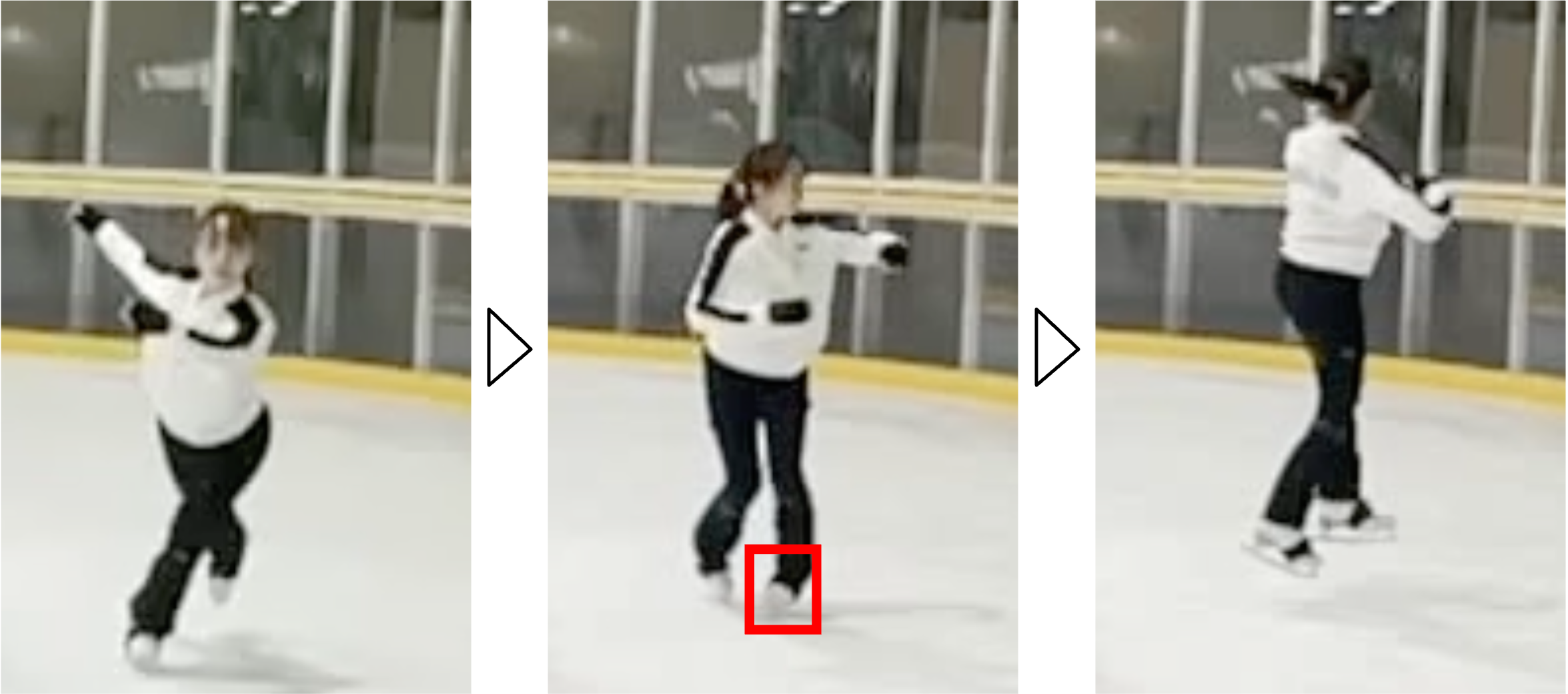}
        \caption{Image examples of take-off with a correct edge (outside edge) from the captured video.}
        \label{fig:success}
    \end{figure}

\begin{figure}[h]
        \centering
        \includegraphics[scale=0.34]
        {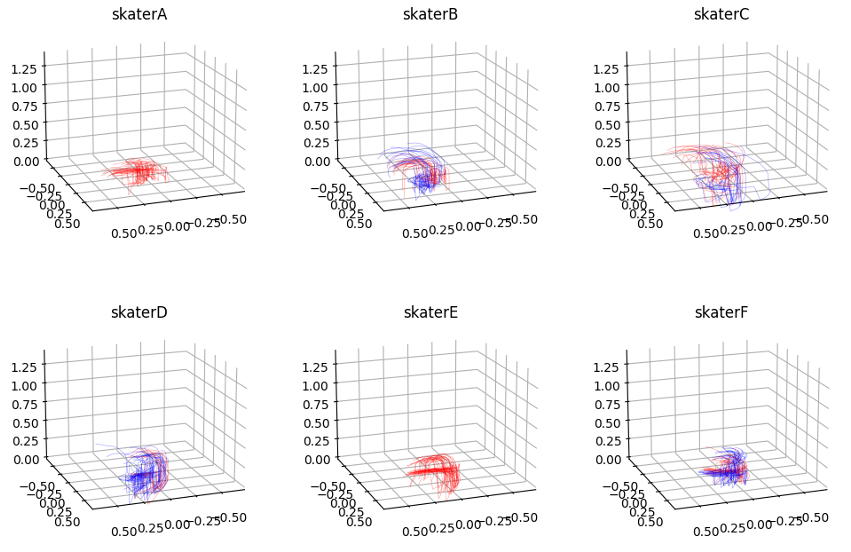}
        \caption{The trajectories of left foot captured by PN3. The configurations are the same as Figure \ref{fig:LF_STpose3D}. The trajectories of the red line almost overlapped the trajectories of the blue line.}
        \label{fig:LF_PN3}
    \end{figure}

For more detailed analysis, Figure \ref{fig:LF_STpose3D} shows the trajectories of estimated positional coordinates of the left foot, which had the highest feature importance in the model using features estimated by ST-Pose3D. Note that the center of the x-y coordinates was normalized to the hip position coordinates. Figure \ref{fig:LF_STpose3D} shows that the blue line drew a straight trajectory while the red line drew a trajectory that bulged outward in a rounded manner in skaters B, C, and F. For example, Figures \ref{fig:error} and \ref{fig:success} show a take-off moment of skater F in videos of error and correct edges, respectively. It could be observed that the left foot was far from the hips at an edge-error jump; by contrast, the left foot was close to the hips at a correct jump. This suggests that the 3D pose estimation by ST-Pose3D could represent this difference in the relative position relationship of each joint. Figure \ref{fig:LF_PN3} shows the trajectories of the left foot position coordinates captured by PN3. The trajectories of the red lines of the edge-error jump almost overlapped the trajectories of the blue lines of the correct jump, and this indicates that IMUs might not be able to capture the differences in left foot position between edge-error jumps and correct edge jumps.

%%%%%%%%%%%%%%%%%%%%%%%%%%%%%%%%%%%%%%%%%%%%%%%%%%%%%%%%%%%%%%%%%%%%%%%%%%%%%%%
%%%%%%%%%%%%%%%%%%%%%%%%%%%%%%%%%%%%%%%%%%%%%%%%%%%%%%%%%%%%%%%%%%%%%%%%%%%%%%%

%%%%%%%%%%%%%%%%%%%%%%%%%%%%%%%%%%%%%%%%%%%%%%%%%%%%%%%%%%%%%%%%%%%%%%%%%%%%%%%
%%%%%%%%%%%%%%%%%%%%%%%%%%%%%%%%%%%%%%%%%%%%%%%%%%%%%%%%%%%%%%%%%%%%%%%%%%%%%%%

%%%%%%%%%%%%%%%%%%%%%%%%%%%%%%%%%%%%%%%%%%%%%%%%%%%%%%%%%%%%%%%%%%%%%%%%%%%%%%%
%%%%%%%%%%%%%%%%%%%%%%%%%%%%%%%%%%%%%%%%%%%%%%%%%%%%%%%%%%%%%%%%%%%%%%%%%%%%%%%

%%%%%%%%%%%%%%%%%%%%%%%%%%%%%%%%%%%%%%%%%%%%%%%%%%%%%%%%%%%%%%%%%%%%%%%%%%%%%%%
%%%%%%%%%%%%%%%%%%%%%%%%%%%%%%%%%%%%%%%%%%%%%%%%%%%%%%%%%%%%%%%%%%%%%%%%%%%%%%%

To quantify the above observations, the average Euclidean distance was calculated for the difference between the trajectories drawn by the correct jump and the edge-error jump, estimated by ST-Pose3D and the inertial sensors. Table \ref{tab:Euclid_dst} shows the results and the standard deviations.

%%%%%%%%%%%%%%%%%%%%%%%%%%%%%%%%%%%%%%%%%%%%%%%%%%%%%%%%%%%%%%%%%%%%%%%%%%%%%%%
    \begin{table}[ht]
        %\vspace{-5pt}
        \caption{The Means of Euclid Distance}
        \centering
        \scalebox{1}{
        \begin{tabular}{lcc}
            \hline
            & \multicolumn{1}{c}{ST-Pose3D} & \multicolumn{1}{c}{PN3} \\
            \hline
            skater B & $2.360 \pm 0.308$ & $2.104 \pm 0.679$\\
            skater C & $1.552 \pm 0.539$ & $2.996 \pm 0.829$\\
            skater D & $1.992 \pm 0.794$ & $1.684 \pm 0.662$\\
            skater F & $1.950 \pm 0.784$ & $1.412 \pm 0.283$\\
            \hline
        \end{tabular}
        }
        \label{tab:Euclid_dst}
        %\vspace{-10pt}
    \end{table}
%%%%%%%%%%%%%%%%%%%%%%%%%%%%%%%%%%%%%%%%%%%%%%%%%%%%%%%%%%%%%%%%%%%%%%%%%%%%%%%

\begin{figure}[h]
        \centering
        \includegraphics[scale=0.55]
        {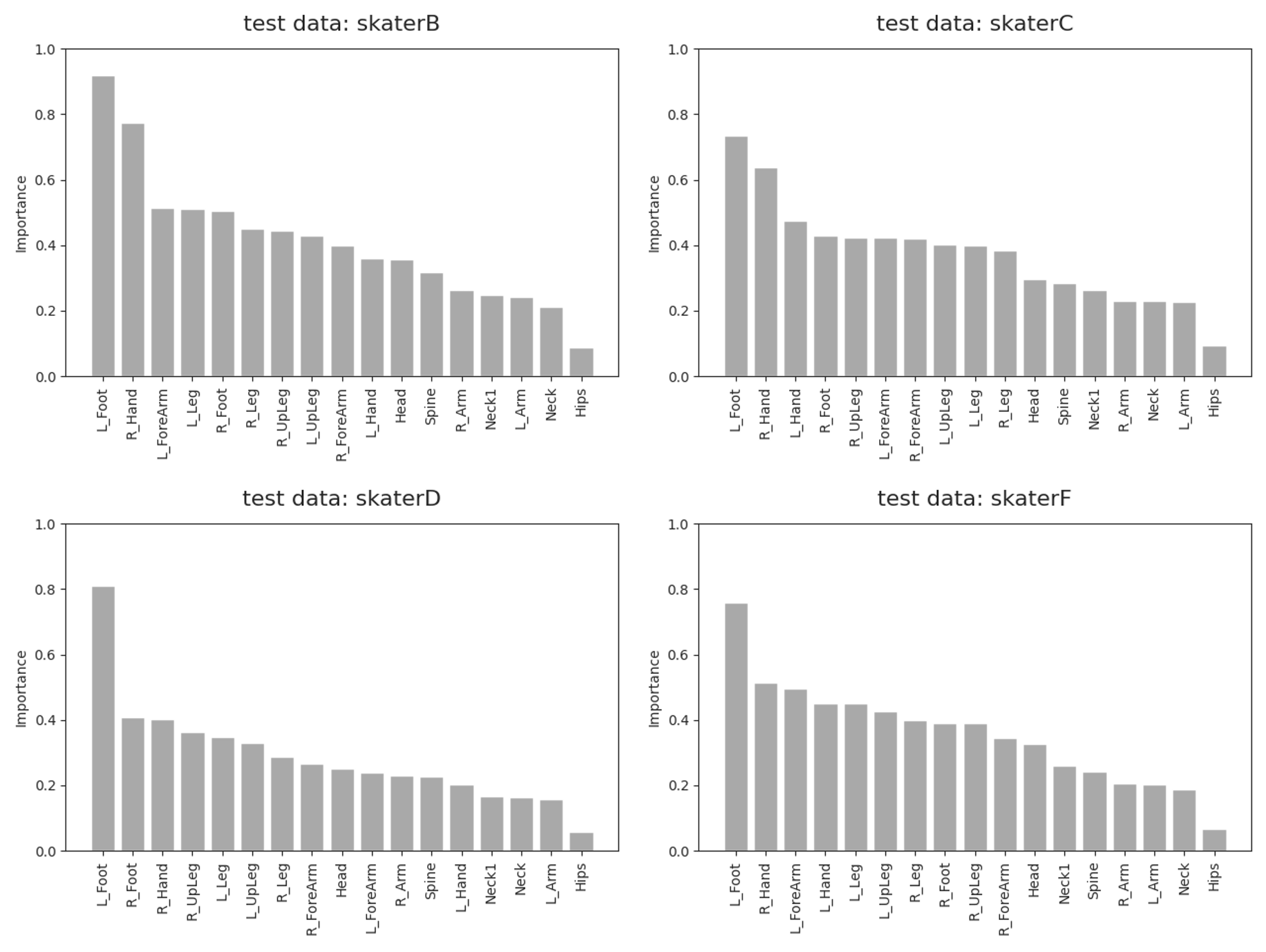}
        \caption{The feature importance (joint pos. 12fps ST-Pose3D) for four skaters, including the correct edges and edge errors. For Skater C, the feature importance of the left foot was not greatly different from the other features.}
        \label{fig:coef_skaters}
    \end{figure}

In the jump of skater C, the Euclidean distance in the inertial sensor was larger than in ST-Pose 3D. However, conversely, for skaters B, C, and D, the Euclidean distance in ST-Pose 3D was larger than that in the inertial sensor. This suggests that the joint position coordinates estimated by ST-Pose3D may be a more practical feature for judging edge errors for the three skaters.
The feature importance of the model using the position coordinates estimated by ST-Pose3D with skaters B, C, D, and F as test data is shown in Figure \ref{fig:coef_skaters}. The feature importance of the left foot did not so much differ from the other features in skater C, for which Euclidean distance was smaller than the other three skaters. This suggests that the difference in Euclidean distance may have affected the feature's importance.

%%%%%%%%%%%%%%%%%%%%%%%%%%%%%%%%%%%%%%%%%%%%%%%%%%%%%%%%%%%%%%%%%%%%%%%%%%%%%%%
%%%%%%%%%%%%%%%%%%%%%%%%%%%%%%%%%%%%%%%%%%%%%%%%%%%%%%%%%%%%%%%%%%%%%%%%%%%%%%%

\begin{figure}[h]
        \centering
        \includegraphics[scale=0.33]
        {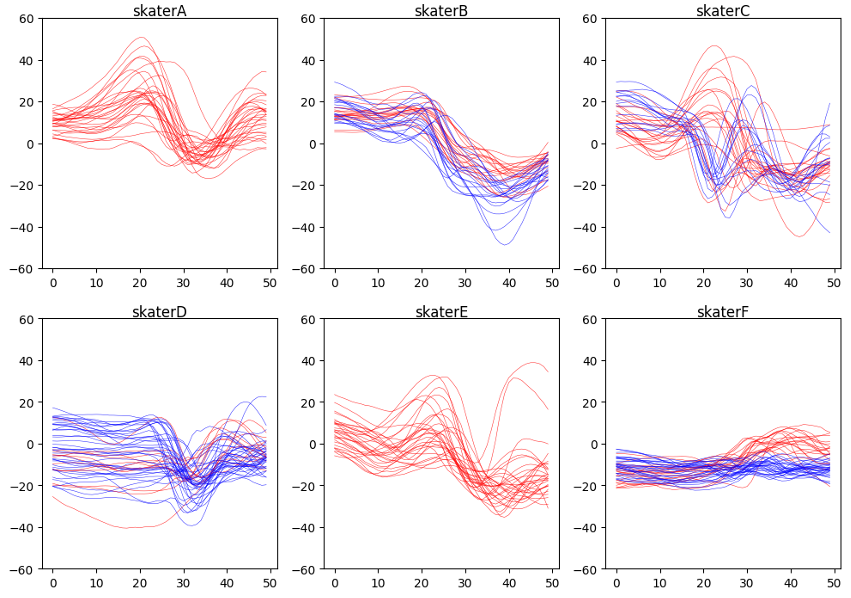}
        \caption{Left ice skate pose angles captured by PN3. The vertical axis is the angle (in degrees), and the horizontal axis is the number of frames. The coloring rule is the same as in Figures \ref{fig:LF_STpose3D} and \ref{fig:LF_PN3}.}
        \label{fig:LF_angle}
    \end{figure}

We also examined the model using the left skate pose angle with the lowest judgment accuracy. Figure \ref{fig:LF_angle} shows the time variation of the pose angle of each skater's left ice skate. The horizontal axis represents the number of frames, and the moment of the jump is approximately the 25th frame. The vertical axis represents the pose angle of the left ice skate, with positive values indicating that the skater used the inside edge and negative values indicating that the skater used the outside edge. For skaters A, B, C, and E, the red lines, judged as an edge error, show an upward convex trend to take a positive value at approximately the 25th frame, the take-off moment.
By contrast, the shape of the graph of the blue line, judged as a correct edge, varied among skaters, but the angle took a negative value from about the 20th frame. This suggests that PN3 can approximately detect the tilt of the ice skate blades. However, for skater F, we found a difference between the actual movement and the sensing results. Figure \ref{fig:error} is an image of the moment when the skater jumped on the error edge, and it can be seen that the left foot leaned far to the inside. However, in Figure \ref{fig:LF_angle}, skater F's left foot (about the 20th frame) was sensed as leaning to the outside by about 0 to 20 degrees. Possible reasons for this discrepancy between actual and sensing results include sensor mounting position, calibration, and estimation error, which would need to be verified in the future when using the IMU. This experiment showed that there was a possibility that there was a discrepancy between the actual and the measured results using the IMUs.

%%%%%%%%%%%%%%%%%%%%%%%%%%%%%%%%%%%%%%%%%%%%%%%%%%%%%%%%%%%%%%%%%%%%%%%%%%%%%%%
%%%%%%%%%%%%%%%%%%%%%%%%%%%%%%%%%%%%%%%%%%%%%%%%%%%%%%%%%%%%%%%%%%%%%%%%%%%%%%%

\section{Conclusion}

% まとめ
In this paper, edge error judgment in figure skating was automated with logistic regression using 3D pose estimation from a monocular camera and IMUs. We compared and evaluated the differences in judgment results depending on the input features and evaluated the processing and measurement methods required for better judgments. As a result, the model with the 3D joint position coordinates from the monocular camera downsampled at 12 fps as input features had the highest accuracy of $83\%$.

% 課題と今後の方針
One of the challenges for the future is first to collect jump data from more skaters because there was a considerable variation in judging among participants. Second, although the highest judgment accuracy model used 3D joint position coordinates as input features, the model was not strictly based on the rule because the rules define edge error in terms of blade inclination. In future work, estimating the ice skate's pose angle through monocular 3D pose estimation would be expected because this study showed that 3D pose estimation from a monocular camera might be practical for judging.

%% identification of the section in the article metadata and the
\begin{acks}
This work was supported by JSPS KAKENHI (Grant Numbers 20H04075 and 21H05300) and JST Presto (Grant Number JPMJPR20CA).
\end{acks}

\bibliographystyle{ACM-Reference-Format}
\bibliography{reference}

\appendix

\end{document}